# Signature of multi-channel interference in high-order harmonic generation from $N_2$ driven by intense mid-infrared pulses

Guihua Li[1,3], Jinping Yao[1], Xinhua Xie[4], Hongqiang Xie[1,3], Bin Zeng[1], Wei Chu[1], Chenrui Jing[1,3], Jielei Ni[1,3], Haisu Zhang[1,3], Xiaojun Liu[2‡], Jing Chen[5,6 †], Ya Cheng[1, *], and Zhizhan Xu[1, #]

[1] State Key Laboratory of High Field Laser Physics, Shanghai Institute of Optics and Fine Mechanics, Chinese Academy of Sciences, Shanghai 201800, China

[2] State Key Laboratory of Magnetic Resonance and Atomic and Molecular Physics, Wuhan Institute of Physics and Mathematics, Chinese Academy of Sciences, Wuhan 430071, China

[3] Graduate School of Chinese Academy of Sciences, Beijing 100080, China

[4] Photonics Institute, Vienna University of Technology, A-1040 Vienna, Austria, EU[4]

[5] HEDPS, Center for Applied Physics and Technology, Peking University, Beijing 100084, China

[6] Institute of Applied Physics and Computational Mathematics, P. O. Box 8009, Beijing 100088, China

‡Corresponding author: xjliu@wipm.ac.cn
†Corresponding author: chen_jing@iapcm.ac.cn
*Corresponding author: ya.cheng@siom.ac.cn
#Corresponding author: zzxu@mail.shcnc.ac.cn

**Abstract** We investigate the multi-electron dynamics in high-order harmonic generation (HHG) from $N_2$ molecules. Clear spectral minima are observed in the cutoff region at all three mid-infrared wavelengths (i.e., 1300, 1400 and 1500 nm) chosen in our experiment. It is found that the positions of the spectral minima do not depend on the alignment angles of molecules. In addition, the spectral minima shift almost linearly with the increasing laser intensity at all three wavelengths, which provides a strong evidence on the dynamic multi-channel interference origin of these minima. The advantages of observation of dynamic multi-channel interference based on HHG driven by long wavelength lasers are discussed.

**PACS** number(s): 42.65.Ky, 42.65.Sf

# 1. Introduction

High-order harmonic generation (HHG) from gas media driven by intense laser fields has been extensively investigated in the past three decades. HHG can be described with a simple three-step model [1,2]: (1) tunneling of electron through the potential barrier formed by distortion of Coulomb potential with strong laser field to continuum; (2) excursion of the free electron in the laser field. Some electrons may be driven back by the oscillating laser field; and (3) recombination of the electron with the parent ion, giving rise to emission of a high-harmonic photon. Owing to its inherent highly nonlinear nature, HHG provides attractive routes to generation of coherent tabletop X-ray sources and attosecond pulses [3-5]. In addition, HHG has also shown a promising future for self-probing of the structures and dynamics in molecules [6-12], as the characteristics of harmonics, i.e., the order-dependent phase and amplitude of harmonics, convey important (both static and dynamic) information of molecular orbitals.

In the early studies of HHG from molecules, it was widely believed that only the highest occupied molecular orbital (HOMO) of molecules should be considered as the major contributor to HHG, as the tunnel ionization scales exponentially to the ionization potential. Recently, clear evidences of the participation of inner orbitals in HHG have been found [13,14], due to the small energy interval between HOMO and the lower-lying orbitals, as well as the geometries of the molecular orbitals. As a consequence, interference between the harmonics originated from different orbitals may occur, leading to formation of minima (dips) in the HHG spectra [10,15,16].

It should be clarified here that to date, three major types of minima have been observed in HHG spectrum. Taking HHG from $N_2$ molecules as an example, first of all, Cooper-like minimum was first discovered near ~40 eV, whose spectral position remains unchanged regardless of the driver laser parameters (e.g., intensity, wavelength, etc.) and molecular alignment angles [17-19]. Secondly, as predicted by Lein [20,21], spectral minimum can also be induced by destructive interference between the harmonics originating from different atoms in a molecule, whose spectral position can be strongly affected by the alignment angles but not the laser parameters at all [22-25]. Thirdly, a dynamic minimum (i.e., multi-channel interference minimum) can originate from the destructive interference between harmonics originating from different orbitals, whose spectral position strongly depends on the laser parameters but not the alignment angles. Identification of the origin of the minimum in a HHG spectrum is critical for self-probing of the molecular structures and dynamics, which has been an important subject in attosecond physics.

In this paper, we focus on the investigation of multi-channel interference in HHG from $N_2$ molecules at three different driver wavelengths, i.e., 1300, 1400 and 1500 nm. It is noteworthy that for the spectral minima of HHG from $N_2$, most previous investigations were focused on low energy region near ~40 eV due to the relatively lower cutoff energy driven by shorter driving wavelengths (e.g., 800 nm) [7,14,17,25,26]. In this low-energy region, different types of minima might be closely located, e.g., the Cooper minimum and structural minimum of $N_2$ are both located near ~40 eV. In addition, it is also difficult to perform detailed investigation of the dynamic minimum in HHG from $N_2$ at ~800 nm wavelength because of the limited photoelectron energy and the sparse harmonic orders. We show that the observation of dynamic minimum in HHG from $N_2$ can be easily realized by using wavelength-tunable ultrafast laser pulses in the mid-infrared (mid-IR) region. In this case, thanks to the wavelength scaling of the HHG energy cutoff [27], the HHG spectra can be dramatically extended toward very high photon energies [28]. In the meantime, the spectral spacing between two neighboring harmonics can be reduced with increasing driving wavelength. These advantages lead to a broader parameter range for observation and investigation of the spectral minimum formed by dynamic multi-channel interference.

## 2. Experimental

Our pump–probe experimental setup is schematically illustrated in Fig. 1, in which the pump pulses are used to align molecules, and the probe pulses to generate high-order harmonics. For HHG experiment with unaligned molecules, the pump pulses are blocked. 6-mJ laser pulses from a Ti:Sapphire laser system (Legend Elite-Duo, Coherent Inc.), which have a central wavelength of 800 nm, a pulse duration (FWHM) of ~40 fs, and a repetition rate of 1 kHz, were split into two parts with a pulse energy ratio of 1:19. The low intensity pulses were used as the pump pulses, whereas the high intensity pulses were used to pump an optical parametric amplifier (OPA, HE-TOPAS, Light Conversion, Ltd.) to generate probe pulses with tunable wavelengths ranging from ~1150 nm to ~2500 nm. In order to efficiently align the $N_2$ molecules, the pump pulses were stretched to ~70 fs before passing through an inverted telescope to reduce beam diameter by a factor of two, and in the meantime, the pump laser intensity was carefully adjusted by controlling the diameter of an aperture in the pump-beam path. The time delay between the pump and probe pulses was fixed at ~4.1 ps, which corresponds to the best alignment (i.e., the $N_2$ molecular axis is parallel to the polarization direction of the pump pulses) at the half-revival of $N_2$ molecular rotation. A half-wave plate was inserted into the pump-beam path to adjust the relative angle between the polarization axes of the pump and probe pulses. Initially, the pump and probe pulses were linearly polarized in the same direction. The pump and probe pulses were recombined and collinearly focused by a position-adjustable lens into the $N_2$ jet with a diameter of 0.5 mm and a back pressure up to 3 bar. The high-order harmonic emission was recorded with a soft-x-ray CCD placed behind a spectrometer [29]. The peak intensity of the probe pulses can be continuously varied by changing the aperture diameter or adjusting the laser power by simply changing the incident angle of a thin glass plate in the light path [30]. To optimize the HHG signal, the positions of lens and gas jet as well as the aperture diameter were adjusted for different measurements.

## 3. Experimental results

Figs. 2 (a-d) show the HHG spectra recorded under the same experimental conditions but at different peak intensities of driver pulses. Here, laser pulses at 1500 nm wavelength were used to generate the high-order harmonics from randomly aligned N$_2$ molecules. The focal spot was positioned ~1.5 cm after the gas jet, giving rise to a favorable phase matching condition for the long-trajectory contribution [31]. Under these conditions, clear minima can be seen in HHG spectra as presented in Figs. 2 (a)-(c). At a peak intensity of $I = 1.97 \times 10^{14}$ W/cm$^2$, a spectral minimum was observed at ~101 eV, which is not far from the cutoff energy ~116 eV (Fig. 2(a)). When the peak intensity was decreased to $1.80 \times 10^{14}$ W/cm$^2$ and $1.63 \times 10^{14}$ W/cm$^2$, the minimum shifted to ~92 eV and ~85 eV in their HHG spectra, respectively ( Figs. 2 (b) and 2(c)). It was also observed that the spectral minimum disappears when the laser intensity was reduced to ~$1.44 \times 10^{14}$ W/cm$^2$ (Fig. 2 (d)). It is noteworthy that in all the above descriptions, the peak intensities of the driver laser fields were evaluated based on the cutoff energies of HHG spectra. Specifically, in consideration of the influence of phase matching on the cutoff [32-35], an expression $E_{cutoff\_exp} \propto I\lambda^{1.7}$ was employed to evaluate the peak intensity of laser field, where $I$ is the peak intensity of driving pulses and $\lambda$ is the laser wavelength.

Furthermore, we investigate the dependence of the observed spectral minima on the alignment angle ($\theta$) between the molecular axis and the polarization direction of probe pulses, as shown in Fig. 3. In our experiments, the best alignment of N$_2$ molecules occurs at ~4.1 ps (i.e., half-revival) after the arrival of the pump pulses, with a typical alignment degree of $\langle \cos^2 \theta \rangle = 0.6$ calculated based on Refs. [36,37]. First, we generate high-order harmonics with the probe pulses whose polarization direction is parallel to the molecular axis at a peak intensity of $2.16 \times 10^{14}$ W/cm$^2$, as shown in Fig. 3(a). Similar to the HHG spectra in Fig. 2, a minimum is again observed around 105 eV, although this spectral dip (i.e., spectral minimum) appears less pronounced. Rotating the polarization of the pump pulses by 90° in the plane perpendicular to the laser propagation direction leads to the HHG spectrum shown in Fig. 3(b). In such case, the probe pulse was polarized perpendicular to the molecular axis. A more pronounced spectral dip is observed at the same photon energy (i. e., ~105 eV) in the HHG spectrum, indicating that the position of spectral dip almost has no dependence on the alignment angle of the molecule.

Lastly, Fig. 4 shows three HHG spectra recorded at different driver wavelengths, i.e., 1500, 1400 and 1300 nm. We carefully adjusted the peak intensities of driving pulses at these wavelengths to ensure that the cutoff energies of all the HHG spectra recorded at these wavelengths could have a similar value, i.e., ~110 eV. We observed that for the 1500 nm driving wavelength, a clear dip appeared at ~92 eV in the vicinity of cutoff (Fig. 4 (a)). Switching to 1400 nm driving wavelength leads to a shift of the dip toward the low-energy end of HHG spectrum, which is at ~86 eV, as shown in Fig. 4 (b). Interestingly, when the driving wavelength was set at ~1300 nm, the spectral dip again appears close to the cutoff, which is at~93 eV as shown in Fig. 4 (d).

## 4. Discussion

In all the experimental investigations mentioned above, we have observed clear spectral minima in harmonic spectra driven by mid-IR laser pulses from $N_2$ molecules. As already mentioned, three types of spectral minima (i.e., Cooper-like minimum, two-center interference minimum and multi-channel interference minimum) have been found in molecular HHG experiments. Identification of the origin of minimum is a crucial step toward the molecular orbital reconstruction. The three types of minima exhibit different dependences on laser parameters as well as molecular alignment angles. For example, the position of multi-channel interference minimum is sensitive to not only laser peak intensity but also driver wavelength, whereas both the Cooper-like minimum and the two-center interference minimum are not. On the other hand, the position of two-center interference minimum can be strongly influenced by the alignment angle of molecules, while the positions of the other two types of minima have no dependence on molecular alignment angle. Therefore, based on our experimental observations, we can safely attribute the spectral minima appeared in our HHG spectra to the mechanism of multi-channel interference, since the position of this minimum changes with the laser peak intensity and the driver wavelength as well. They are, however, not sensitive to the alignment angle of the molecule. All of these features are consistent to that of multi-channel interference minimum. In addition, due to alignment-angle dependence of the relative contribution of HOMO and HOMO-1 to HHG [7,14], the depth of spectral dip originated from multi-channel interference should be sensitive to the alignment angle of N$_2$ molecule, while the Cooper-like and two-center interference minima do not exhibit such behavior. In our pump-probe experiment, the spectral dip appears more pronounced when the molecule is aligned perpendicular to the polarization of probe pulses than that of the parallel case (Fig. 3), which provides further evidence on the mechanism of multi-channel interference.

Furthermore, we have found that in our HHG experiments carried out at all the three wavelengths, the position of spectral minima shows a linear dependence on the laser peak intensity, as evidenced in Fig. 5. This finding motivates us to make a more thorough theoretical analysis on our experimental results as follow.

In the simplest form, the position of the minimum observed in our experiment can be reproduced with an expression as follow [7,10,24,38]:

$$\Delta \varphi = \Delta I_p \times \tau + \phi = (2n+1)\pi, n = 0,1,2,3\cdots \qquad (1)$$

Here, $\Delta \varphi$ is the phase difference between the harmonics generated from the two channels (i.e., contributed by electrons ionized from HOMO and HOMO-1 orbitals), $\tau$ is the electron excursion time in the continuum state, $\Delta I_p$ is the difference between the ionization potentials of two orbital, i.e., $\Delta I_p = I_{p(HOMO-1)} - I_{p(HOMO)} = 1.4eV$ and $\phi$ is an additional phase which is artificially chosen for achieving best fitting of the experimental data. Based on Eq. (1), we theoretically calculated the positions of dynamic minima by using a classical three-step model [1] and found out the best fitting $\phi$ at a given driving wavelength. Specifically, we find that, for 1300 nm and 1400 nm wavelengths, the position of minima can be well reproduced with the expression $\Delta \varphi = \Delta I_p \times \tau - 0.5\pi = \pi$, where $\tau = 2.23$ fs is constant for different peak laser intensities; whereas for 1500 nm wavelength, the expression for fitting the experimentally measured positions of minima should be changed to $\Delta\varphi = \Delta I_p \times \tau + 0.5\pi = 3\pi$, where $\tau = 3.83$ fs. For comparison, the experimentally measured and theoretically calculated positions of minima at all the three wavelengths are plotted as functions of the cutoff energies in Fig. 6. Perfect linearity and good agreements between the experimental and theoretical results are obtained in all the cases.

It is noteworthy that in the above calculations, an additional phase difference $\phi = \pm 0.5\pi$ is always needed for achieving best fitting results, whose origin may be attributed to the ionization and/or recombination processes in the HHG. According to Ref. [9], the influence of ionization process on the phase difference might be ignored since the photoionization occurs in the deep tunnelling regime (i.e., Keldysh parameter < 1) under our experimental condition. On the other hand, the phase difference of $\varphi = \pm 0.5\pi$ appears due to different parities of participating orbits in the recombination process [7,10,24]. More recently, Z. Diveki *et al* have reported that for multi-channel interference, there could be an additional difference in the recombination phases for electrons ionized from different molecular orbitals due to the nuclear dynamics [26]. The role of this additional phase in the multi-channel interference for HHG driven by long wavelength lasers will be investigated in the future work.

## 5. Conclusion

In conclusion, we have experimentally observed dynamic minima near the cutoff of the HHG spectra from $N_2$ molecules driven by intense femtosecond lasers at three wavelengths (i.e., 1300, 1400 and 1500 nm). A systematic investigation has been carried out for clarifying the origin of these minima. We find a linear dependence of the spectral dip position on the peak intensity of the driven field. In addition, the positions of the spectral dips strongly depend on the wavelengths of the driven field, but exhibit no dependence on the alignment angles of the molecule. The experimental results strongly support the multi-channel interference picture rather than the two-center inference picture or Cooper-like minimum picture. Moreover, it should be emphasized that our results demonstrate several significant advantages of using long wavelength driving lasers for investigating dynamic minimum with HHG spectroscopy, which originates from the multi-channel interference in HHG process. First of all, with the long wavelength driving laser, the cutoff of HHG spectrum can be greatly extended, providing broader spectral ranges for occurrence of the interference minimum. Secondly, the denser HHG peaks with smaller spectral spacing at longer wavelengths allow more precise determination of the spectral positions of minima. Thirdly, the wavelength tunability of OPA provides a great flexibility in controlling the minimum position, e. g., making it possible to appear either in cutoff or in plateau regions. Lastly, the small Keldysh parameters at longer wavelengths allow for observation of dynamic minima from complex molecules with relatively low ionization potentials. Thus, the long wavelength driver lasers are attractive for not only generating coherent XUV radiation and attosecond pulses, but also investigating structures and dynamics of molecules in strong laser fields.

This work is supported by the National Basic Research Program of China (Grant Nos. 2011CB808102 and 2013CB9222001), and National Natural Science Foundation of China (Grant Nos. 11127901, 11134010，60921004, 11074026, 10925420, 61275205, 11204332, and 11274050).

**Figure 1:**

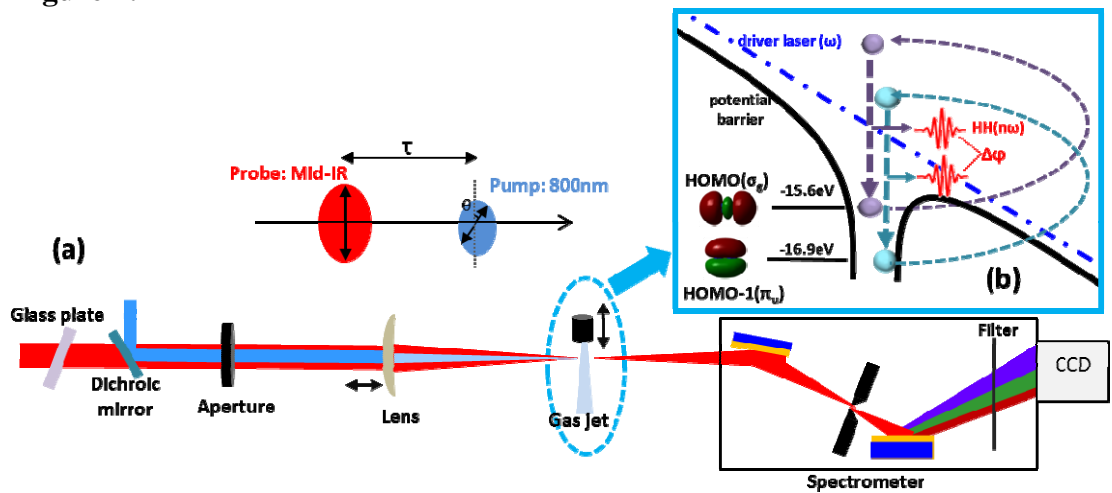

Fig. 1 (Color online) (a) Schematic diagram of the pump-probe experimental setup. (b) The interference process between two different channels from HOMO and HOMO-1 in HHG from $N_2$ molecule.

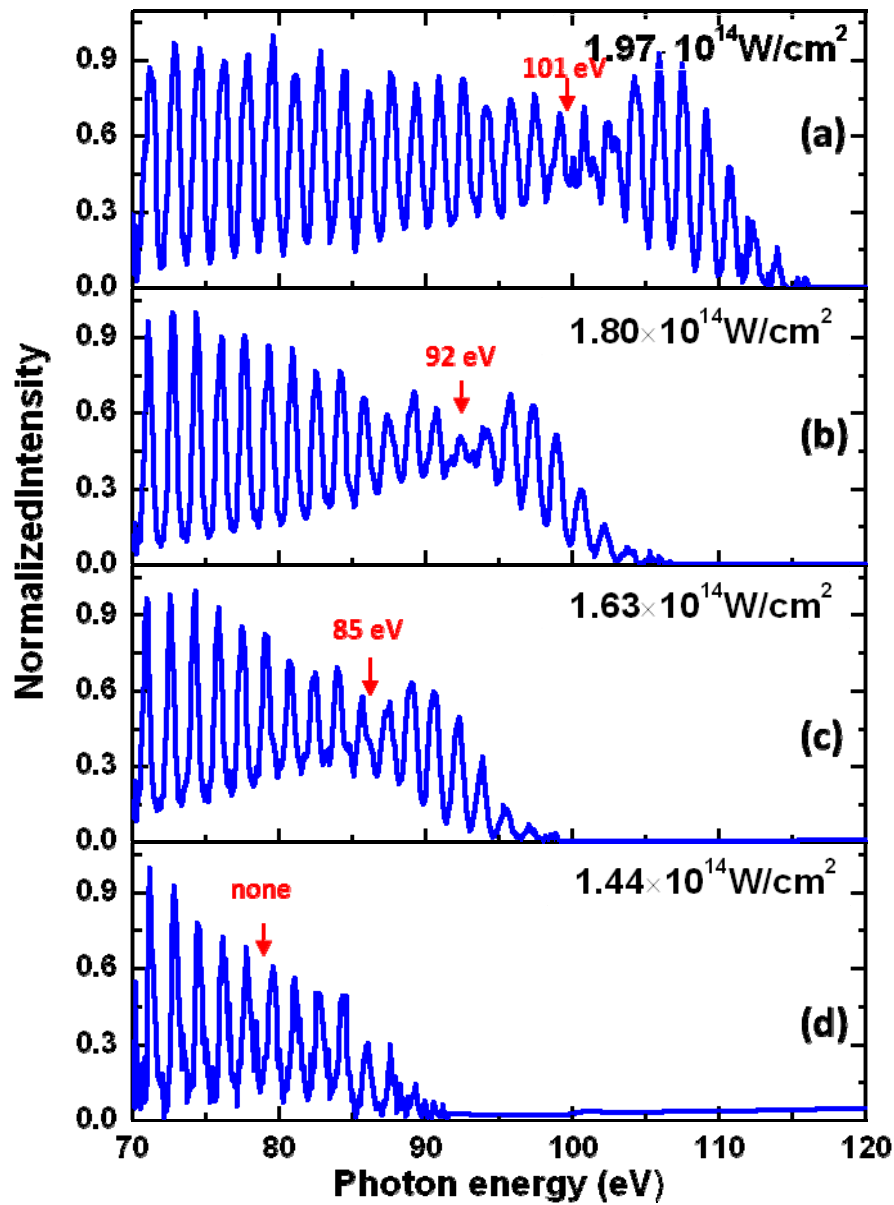

Fig. 2 (Color online) Typical HHG spectra recorded with 1500 nm pulses from unaligned $N_2$ molecules at the laser intensities of (a) $1.97 \times 10^{14}$ W/cm$^2$, (b) $1.80 \times 10^{14}$ W/cm$^2$, (c) $1.63 \times 10^{14}$ W/cm$^2$ and (d) $1.44 \times 10^{14}$ W/cm$^2$.

**Figure 3:**

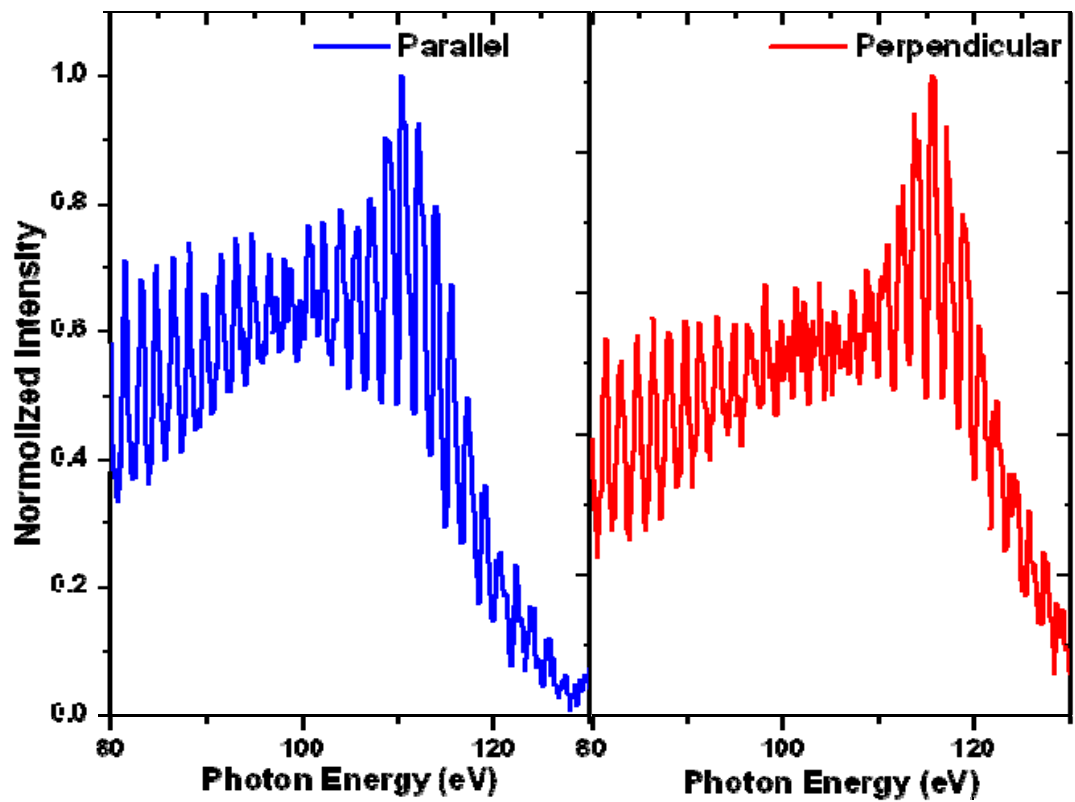

Fig. 3 (Color online) Normalized HHG spectra recorded with 1500 nm laser at a peak intensity of ~2.16×$10^{14}$ W/cm² when $N_2$ molecules are aligned to be (a) parallel and (b) perpendicular to the polarization of probe pulses.

**Figure 4:**

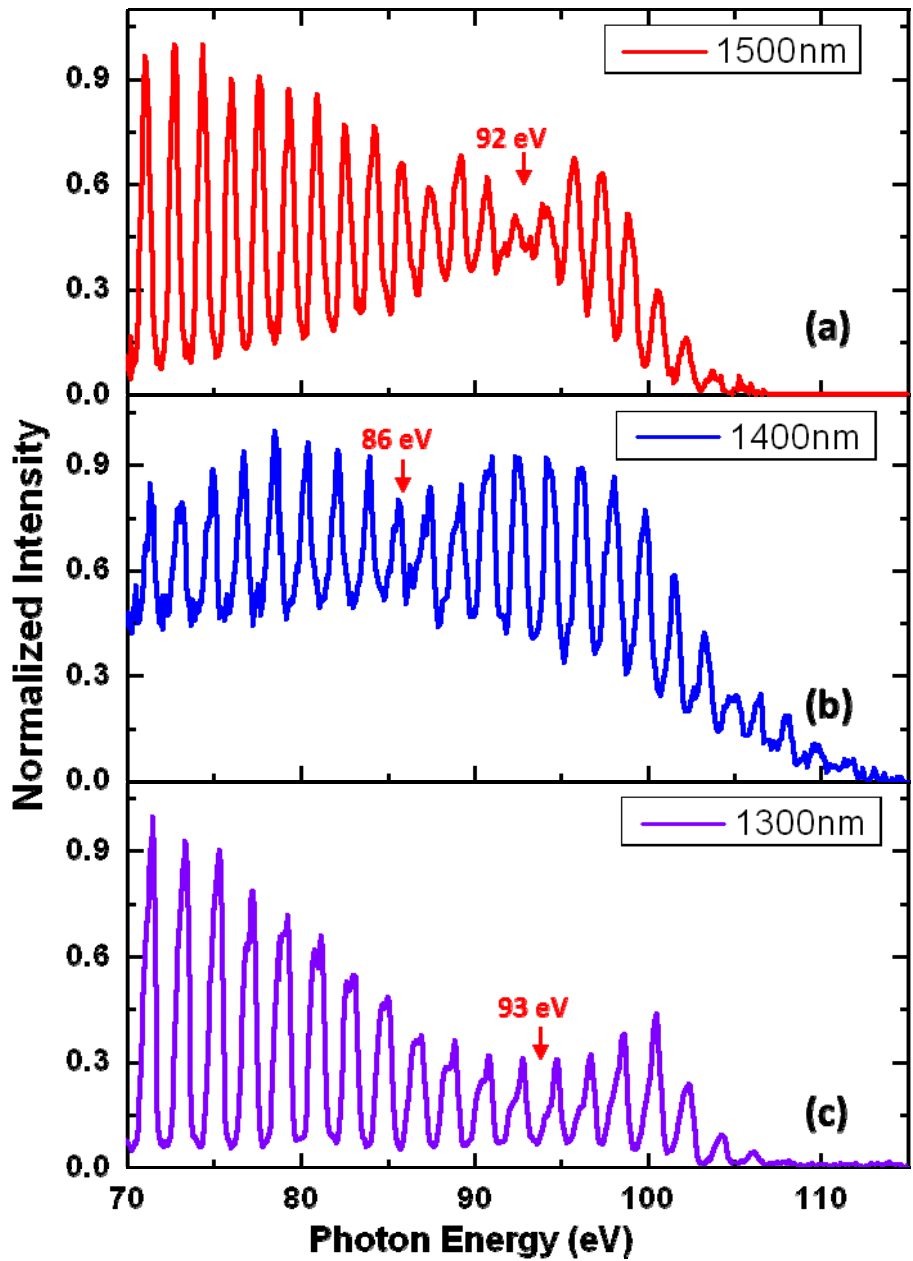

Fig. 4 (Color online) High-order harmonic generated with (a) 1500 nm, (b) 1400 nm and (c) 1300 nm laser pulses from $N_2$ molecules with random alignment. The laser intensities at three wavelengths are $1.80\times10^{14}$, $2.01\times10^{14}$ and $2.30\times10^{14}$ W/cm$^2$, respectively.

**Figure 5**

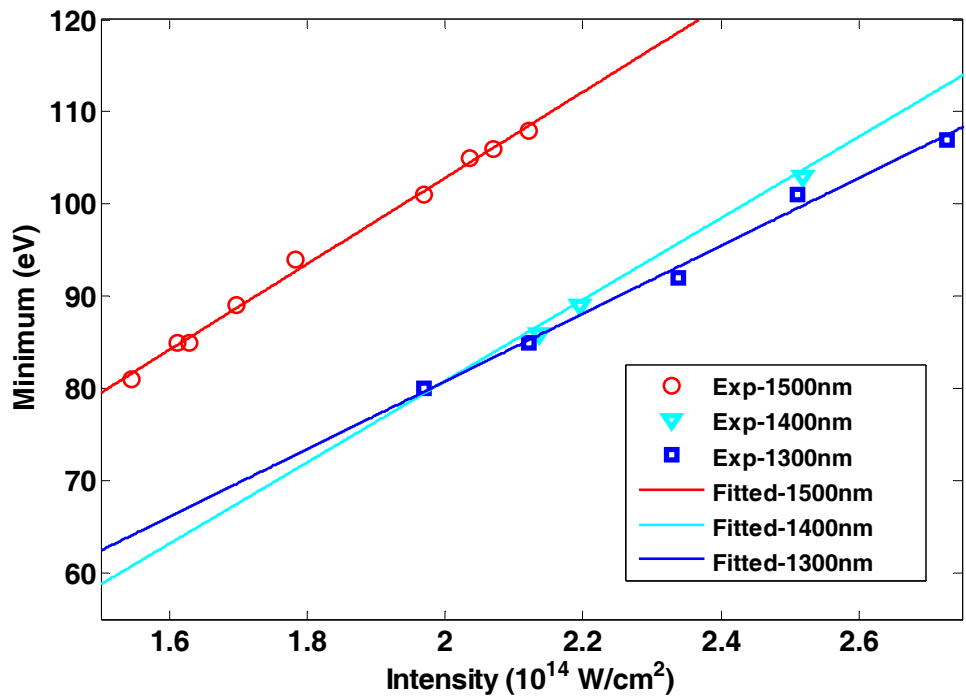

Fig. 5 (Color online) The positions of the spectral minima shift as a linear function of the peak intensity of laser pulses at 1500, 1400 and 1300 nm wavelengths.

**Figure 6**

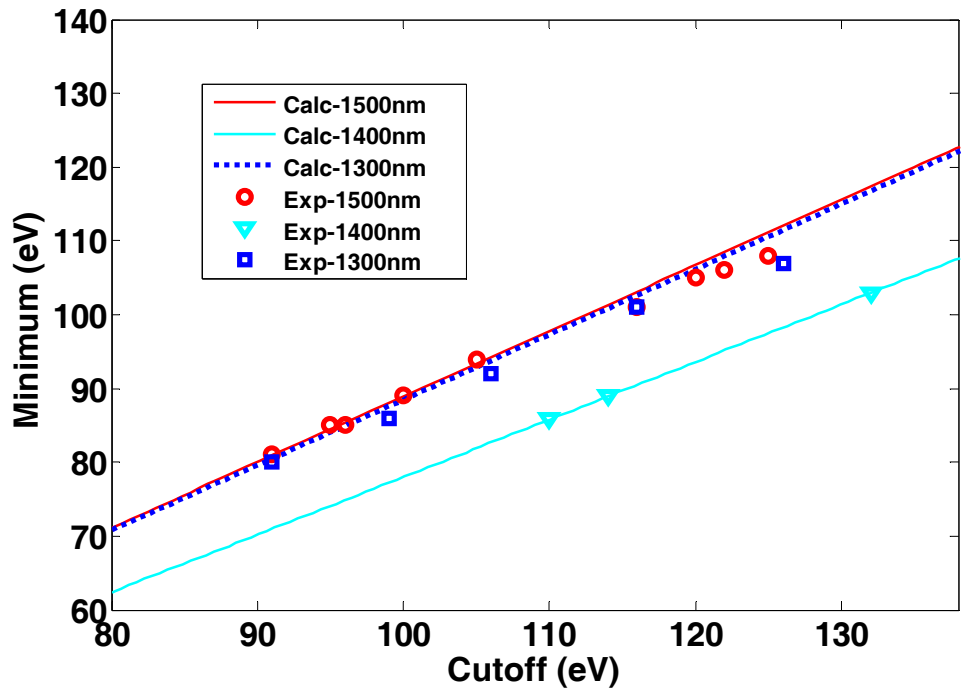

Fig. 6 (Color online) The measured and calculated dip positions in harmonic spectra as a function of cutoff energy for three different laser wavelengths.